# Performance comparison of IEEE 802.11g and IEEE 802.11n in the presence of interference from 802.15.4 networks

S. Haani Masood
Department of Electrical Engineering,
McGill University
Email: syed.h.masood@mail.mcgill.ca

*Abstract* - *Recent advances in wireless technology has led to the introduction of new devices utilizing the 2.4GHz industrial scientific and medical (ISM) unlicensed band traditionally used by Wireless LANS (WLAN). The most popular amongst them is the IEEE 802.15.4 used in low rate wireless personnel area networks. Moreover, the increasing demand of higher data rate in WLANs has prompted the emergence of the 802.11n protocol which is being widely adopted due to its increased performance (higher data rates up to 300Mbps/channel, MIMO). IEEE 802.11n uses two 20MHz wide channels for its operation, rather than a single 20MHz as in other IEEE 802.11 PHY. Avoiding channel overlap between IEEE 802.15.4 and IEEE 802.11 networks is therefore difficult. Interoperability and coexistence between these networks become key issues and must be catered to guarantee satisfactory performance of both networks. In this paper we compare the packet error rate (PER) and maximum throughput of IEEE 802.11n and IEEE 802.11g under interference from IEEE 802.15.4 by using MATLAB to simulate the IEEE PHY for 802.11n and 802.11g networks.*

## 1. Introduction

IEEE 802.15.4 is establishing its place in the market as an enabler for emerging wireless sensor networks (WSNs) [1]. It utilizes the same 2.4 GHz ISM band as IEEE 802.11 WLAN networks. Due to supporting same complimentary applications, they are likely to be collocated within the interfering range of each other. WLANs on the other hand are striving to achieve the increasing higher data rate demand and its performance under the interference from such networks needs to be evaluated.

There have been some previous studies about the co-existence of IEEE 802.11 with IEEE 802.15.4. According to [2] [3] IEEE 802.11 has a deteriorating affect on the operation of 802.15.4 depending upon channel overlap, however, they do not provide results on the performance of the IEEE 802.11n in such an environment. In [4] the impact of IEEE 802.15.4 on the operation of IEEE 802.11g is investigated under different scenarios using two sensor nodes. Such a study, however, is dependent on the many variables including transmit power of the WLAN and sensor nodes, type of equipment, environment and cannot be generalized. There have been other studies investigating the co-existence of IEEE 802.11b/g networks with IEEE 802.15.4 nodes. To the best of the author's knowledge, there has been no such study for the performance of IEEE 802.11n networks under interference from IEEE 802.15.4.

In this paper, we compare the performance of IEEE 802.11n and IEEE 802.11g in the presence of interference from IEEE 802.15.4 networks. The PER and Maximum Throughput are used as performance measures. The PER is obtained from the Bit Error Rate (BER). Bit Error Rate in such networks is dependent on the Signal to interference and Noise ratio (SINR). The maximum throughput is obtained by measuring the number of successful transmissions of the packets on each network using simulations. To determine these performance metrics we simulate the physical layer of each WLAN protocol (IEEE 802.11n and IEEE 802.11g) in the presence of Flat Fading channels (which is a general assumption in indoor environments [5]). The BER and maximum throughput measurements can be obtained from the results of this simulation. To include the constraints on these measurements provided by IEEE 802.15.4 we next introduce the communications happening over IEEE 802.15.4 as interference for the IEEE 802.11 network. Changes have been made in the physical layer of IEEE 802.11n to support Multiple Input Multiple Output (MIMO-OFDM) communications and increased data-rate (300Mbit/s/channel) which suggests a difference in performance compared to IEEE 802.11g which has a Single Input Single Output (SISO) based physical layer employing only OFDM (54Mbits/s/channel). The analytic results for both protocols are compared with the simulation results.

This paper is organized as follows. Section 2 briefly overviews the IEEE 802.15.4 and IEEE 802.11 protocols. In Section 3, the BER of the IEEE 802.11g and IEEE 802.11n using SINR is evaluated. It also describes the interference model of IEEE 802.15.4 and IEEE 802.11g/n. In Section 4 comparisons between analytic and simulation results are shown. Finally, this paper concludes in Section 5.

## 2. Protocol Overview

### 2.2 IEEE 802.15.4

The IEEE standard, 802.15.4, defines both the physical layer (PHY) and medium access control (MAC) sub-layer specifications for low-rate wireless personal area networks (LR-WPANs), which support simple devices that consume minimal power and typically operate in the personal operating space of 10m or less [1]. Two types of topologies are supported in the IEEE 802.15.4: a one-hop star or multi-hop peer-to-peer topology. The network and upper layers are defined by the ZigBee Alliance [6].

The standard offers two PHY options based on the frequency band. Both are based on direct sequence spread spectrum (DSSS). The data rate is 250 kbps at 868 MHz with binary phase shift keying (BPSK). There is a single channel between 868 and 868.8 MHz, 10 channels between 902.0 and 928.0 MHz, and 16 channels between 2.4 and 2.48435 GHz. Only 3 channels in the 2.4 GHz band can be used for IEEE 802.15.4 which is free of interference from IEEE 802.11 networks. This can be seen from Fig.1. As shown in Fig.1 the channels have a 3MHz bandwidth and are uniformly distributed within the ISM band. Receiver sensitivities are -85 dBm for 2.4 GHz and -92 dBm for 868/915 MHz.

An IEEE 802.15.4 network can work in either beacon-enabled mode or non-beacon-enabled mode. In beacon-enabled mode, a coordinator broadcasts beacons periodically to synchronize the attached devices. In non-beacon-enabled mode, a coordinator does not broadcast beacons periodically, but may unicast a beacon to a device that is soliciting beacons.

For channel access, IEEE 802.15.4 uses slotted/unslotted Carrier Sense Multiple Access with Collision Avoidance (CSMA-CA) mechanism. If an IEEE 802.15.4 station has data to send, it perform random backoff. The backoff window is based on a random value uniformly distributed in the interval [$CW_{min}$,$CW_{max}$], where $CW_{min}$ and $CW_{max}$ represent the Contention Window parameters. After finishing the backoff, the IEEE 802.15.4 station checks the medium using clear channel assessment (CCA). If the medium is sensed idle, it sends its frame. Upon the successful reception of a fram, the destination station return an ACK frame after a short interframe space (SIFS). If the medium is determined busy during CCA period, it doubles the CW size and repeats the basic access procedures. A schematic representation of this mechanism is shown in Fig. 2.

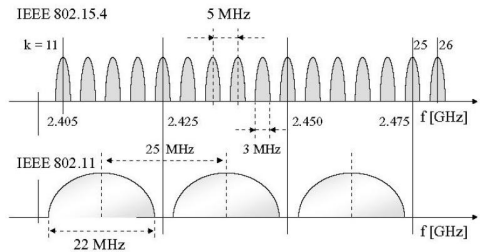

Fig. 1. Frequency channels of IEEE 802.115.4 compared with IEEE 802.11.

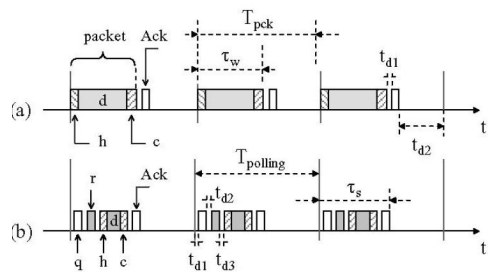

Fig. 2. Time diagram of IEEE 802.11 (a) and IEEE 802.15.4 (b) frames

| Standard | Frequency | Data Rate | Modulation | MIMO? | Range |
|---|---|---|---|---|---|
| 802.15.4 | 2.4GHz | 250Kbps | DSSS | No | 10m |
| 802.11g | 2.4GHz | 6 Mbps | OFDM (BPSK) | No | 140m |
| 802.11n | 2.4GHz | 30 Mbps | OFDM (BPSK) | Yes | 250m |

Table 1. Comparison of different wireless standards

### 2.3 IEEE 802.11g

The IEEE 802.11 standard defines both the physical (PHY) and medium access control (MAC) layer protocols for WLANs [7]. The standard operates in a total of 14 channels available in the 2.4-GHz band, numbered 1 to 14, each with a bandwidth of 22MHz and a channel separation of 5MHz. This channel mapping can be seen in figure 3. The figure shows that channels are partially overlapped, and that only three channels at a time, e.g. 1, 6 and 11, are not overlapped and can be used without interference between APs. WLAN output powers are typically around 20dBm and operate within a 100m range. The channel contention scheme in MAC layer used by all IEEE 802.11 protocols (IEEE 802.11b/g/n) is the same CSMA/CA as that used in IEEE 802.15.4 which was described earlier. This is depicted in Fig.2.

IEEE 802.11g amendment to the IEEE 802.11 standard was ratified as a third modulation standard in June, 2003 [8]. IEEE 802.11g uses an additional OFDM transmission scheme which was absent in the initial IEEE 802.11 standard. This allows an increase in the maximum physical layer bit rate to up to 54 Mbits/s from 11Mbit/s used by the earlier 802.11b standard utilizing DSSS modulation. The maximum physical layer bit rate achieved in IEEE 802.11g for BPSK modulation is 6Mbits/s/channel. IEEE 802.11g standard is based on Single Input Single Output systems. The modulation scheme based on OFDM is summarized in Table 2.

| Parameter | IEEE 802.11g OFDM | IEEE 802.11n OFDM |
|---|---|---|
| FFT size. nFFT | 64 | 128 |
| No. of sub-carriers | 52 | 114 |
| FFT frequency | 20MHz | 20 / 40 MHz |
| Sub-carrier index | {-26 to -1, +1 to +26} | { -57 to -1, +1 to +57} |
| Cyclic Prefix duration | .8μs | .8μs |

Table 2. IEEE 802.11g and IEEE 802.11n modulation characteristics

### 2.4 IEEE 802.11n

IEEE 802.11n is an amendment to the IEEE 802.11 standard and proposes enhancements for Higher for higher throughput using MIMO-OFDM [9]. IEEE 802.11n brings many new features to deliver the performance gains. It uses the multiple-input-multiple-output (MIMO) technology that enables spatial diversity and spatial multiplexing for respectively increasing the range and data transmission rate. In addition, 802.11n allows use of wider 40 MHz channels to double the bandwidth as compared to the legacy 20 MHz operation. The extension channel (40 MHz) would be used if the existing traffic load on an IEEE 802.11n network cannot be carried within the 20MHz channel. 802.11n uses frame aggregation and block acknowledgements for improving the throughput efficiency. The max physical layer bit rate achievable in IEEE 802.11n is 300Mbits/s/channel using 2 spatial streams on a 40MHz channel. Spatial diversity is achieved by using multiple antennas. The specification allows up to 4 spatial data streams which can implement Space Time Block Code (STBC) schemes as seen in Fig 6.

IEEE 802.11n uses a more efficient OFDM modulation. This more than doubles the data rate for 802.11n when compared to 20 MHz channels. When operating within a traditional 20 MHz channel, OFDM slices the channel into 52 subcarriers (48 of which are used for carrying data). However, when 802.11n applies OFDM on a 40 MHz channel, the number of data-carrying subcarriers increases to 114 subcarriers. This allows 802.11n to deliver a 65 Mbps data rate (instead of 54 Mbits/s) per 20 MHz channel for a total of 135 Mbits/s on a 40 MHz channel when transmitting a single spatial data stream. When transmitting using 2 spatial streams on a 40 MHz channel, this data rate again doubles to 135 Mbits/s x 2 = 270 Mbps. For BPSK type modulation, the maximum physical layer data rate achieved by 802.11n is 30 Mbits/s.

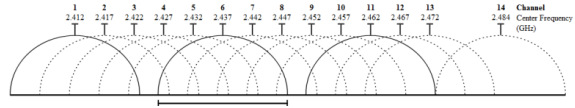

Fig. 3. IEEE 802.11 Channel Assignments

### 2.5 Coexistence of IEEE 802.11 and IEEE 802.15.4

Fig 4. Illustrates an example of coverage overlap of one WLAN and one IEEE 802.15.4 network. In this scenario, devices of each network can experience an interference power from the transmissions of the other network and vice versa. This mutual interference degrades the performance of both the WLAN and IEEE 802.15.4 network.

The relationship between the WLAN (non-overlapping sets) and IEEE 802.15.4 channels at 2.4 GHz is illustrated in Fig 5. To prevent interference between IEEE 802.15.4 and WLAN, IEEE 802.15.4 standard recommends to use the channels that fall in the guard bands between two adjacent non-overlapping WLAN channels or above these channels (this holds for the assumption that most WLAN networks are deployed on non-overlapping channel 1, 6 and 11). While the energy in this guard space is not zero, it is significantly lower than the energy within the channels and operating IEEE 802.15.4 network on one of these channels can minimize interference between systems.

In IEEE 802.11n operating on a 40MHz wide channel this scheme does not work as clearly two or more networks operating in the same location as an IEEE 802.15.4 network would leave no IEEE 802.15.4 channel free from the presence of interference from IEEE 802.11n traffic.

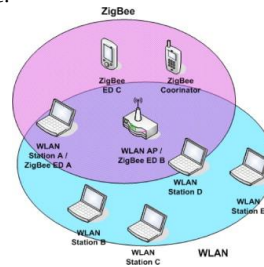

Fig. 4. Coverage overlap example of WLAN and IEEE 802.15.4 network

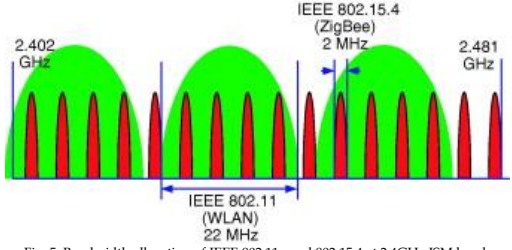

Fig. 5. Bandwidth allocation of IEEE 802.11g and 802.15.4 at 2.4GHz ISM band

## 3. Bit error rate evaluation of IEEE 802.11n and IEEE 802.11g in presence of 802.15.4

In an OFDM transmission, we know that the transmission of cyclic prefix does not carry 'extra' information in Additive White Gaussian Noise (AWGN) channel. The signal energy is spread over time $T_d$ (data symbol duration) and $T_{cp}$ (cyclic prefix duration). Thus, symbol energy is given by:

$$E_s = E_b \cdot T_d / (T_d + T_{CP}) \quad (1)$$

$E_s$ = Symbol Energy
$E_b$ = Bit Energy

In OFDM transmission, all the available subcarriers from the DFT is not used for data transmission. Typically some subcarriers at the edge are left unused to ensure spectrum roll off. For the example scenario, out of the available bandwidth from -10MHz to +10MHz, only subcarriers from -8.1250MHz (-26/64x20MHz) to +8.1250MHz (+26/64x20MHz) are used.

This means that the signal energy is spread over a bandwidth of 16.250MHz, whereas noise is spread over bandwidth of 20MHz (-10MHz to +10MHz).

$$20MHz \times E_s = 16.25MHz \times E_b$$

Simplifying,

$$E_s = \left(\frac{nDSC}{nFFT}\right) \times E_b \quad (2)$$

Combining the above two aspects, the relation between **symbol energy $E_s$** and the **bit energy $E_b$** is as follows:

$$\frac{E_s}{N_0} dB = \frac{E_b}{N_0} dB + 10\log_{10}\left(\frac{nDSC}{nFFT}\right) + 10\log_{10}\left(\frac{T_d}{T_d + T_{CP}}\right) \quad (3)$$

### 3.2 IEEE 802.11g

The probability of bit-error for BPSK is given by:

$$P_{b,BPSK} = \left(\frac{1}{2}\right) \times erfc\left(\sqrt{\frac{E_b}{N_0}}\right) \quad (4)$$

$$P_{S,BPSK} = \left(\frac{1}{2}\right) \times erfc\left(\sqrt{\frac{E_S}{N_0}}\right) \quad (5)$$

### 3.3 IEEE 802.11n

We model our IEEE 802.11n physical layer as consisting of two transmit and two receive antennas (2x2) using OFDM modulation and Space Time Block Coding (STBC). The received signal in Alamouti scheme for 2 transmit and 2 receive antennas as given in [11] is:

$$r = \left((\alpha_1^2 + \alpha_2^2 + \alpha_3^2 + \alpha_4^2) \times s_0\right) + \alpha_1^* n_1 + n_2^* \alpha_2 + \alpha_3^* n_3 + n_4^* \alpha_4$$

From which expression for $E_b/N_0$ can be deduced as:

$$\frac{E_b}{N_0} = \frac{\left((\alpha_1^2 + \alpha_2^2 + \alpha_3^2 + \alpha_4^2) \times s_0\right)^2}{\left((\alpha_1^2 + \alpha_2^2 + \alpha_3^2 + \alpha_4^2) \times var(N_0)\right)}$$

$$\frac{E_b}{N_0} = \frac{\left((\alpha_1^2 + \alpha_2^2 + \alpha_3^2 + \alpha_4^2) \times s_0\right)}{(var(N_0))} \quad (6)$$

$\alpha_n$ : **Channel coefficients n=1, 2, 3, 4 for 2 transmit and 2 receive antennas**
$n_i$ : **AWGN for each channel (i=1,2,3,4)**

Replacing this expression below yields the bit-error probability for IEEE 802.11n:

$$P_{b,BPSK} = \left(\frac{1}{2}\right) \times erfc\left(\sqrt{\frac{E_b}{N_0}}\right) \quad (7)$$

### 3.4 Interference

The signal of the IEEE 802.15.4 can be modeled as a partial band jammer to the WLAN signal and can be approximated as band limited AWGN to the IEEE 802.11g/n network. Then, the SINR can be determined by:

$$SINR = 10\log_{10}\left(\frac{P_c}{P_N + P_i}\right) \quad (8)$$

Where $P_c$, $P_N$ and $P_i$ denote the power of the desired signal, the noise power and the interferer (IEEE 802.15.4 signal) power. This SINR value can be directly replaced into $E_b/N_0$ in the previous equation to get the bit-error rate (BER) for IEEE 802.11n and IEEE 802.11g.

In this paper we consider two different scenarios. In the first scenario we model our IEEE 802.15.4 traffic as interference to both IEEE 802.11g and IEEE 802.11n occurring at the centre frequency of the WLAN. In the second scenario we model the case where only IEEE 802.11n faces interference from IEEE 802.15.4 while IEEE 802.11g is interference free. This case occurs when the centre frequency of our IEEE 802.15.4 network is located in one of the channels which do not overlap with the channels of IEEE 802.15.4 but overlaps with IEEE 802.11n due to its extended 40MHz channel.

## 4. Performance analysis of IEEE 802.11g and IEEE 802.11n in the presence of interference from IEEE 802.15.4

### 4.1 Experiment setup

To simulate the IEEE 802.11g and IEEE 802.11n network we model the physical layer characteristics of both standards in MATLAB as outlined in Table 2 and simulate traffic as an arriving stream of bits modulated as BPSK. In IEEE 802.11n we further implement STBC for 2 transmit and 2 receive antennas for spatial diversity. The scheme used for IEEE 802.11n is depicted in Fig. 6.

Interference from IEEE 802.15.4 is modeled as an increase in AWGN noise occurring at intervals of $\frac{6 \times 10^6}{250 \times 10^3}$ bits of traffic of IEEE 802.11g and $\frac{30 \times 10^6}{250 \times 10^3}$ bits for IEEE 802.11n. This assumption holds if the IEEE 802.15.4 is transmitting at full rate of 250Kbps and we ignore MAC layer channel contention schemes.

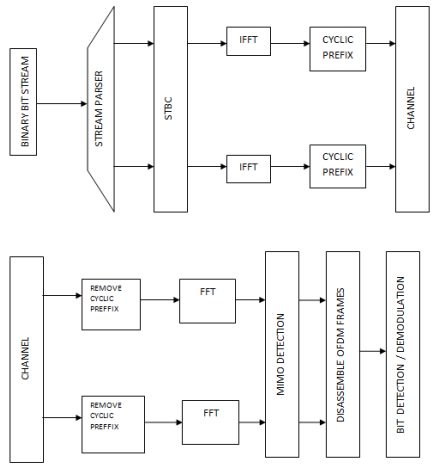

Fig. 6. IEEE 802.11n physical layer implementation (a) Transmitter (b) Receiver

### 4.2 Simulation Results

The performance of IEEE 802.11g vs IEEE 802.11n in the absence of interference from IEEE 802.15.4 is shown in Fig. 6. The simulation results are in accordance with the intuition that MIMO should perform better than SISO due to its higher SNR calculated in equation (6).

The performance of IEEE 802.11n (MIMO-OFDM) is compared with IEEE 802.11g (OFDM) in the presence of interference from IEEE 802.15.4 in Fig. 7. From the plot we can see that the interference provided by the IEEE 802.15.4 is more prominent in IEEE 802.11g as compared to IEEE 802.11n.

The maximum throughput achieved by IEEE 802.11n in the presence of interference is shown in Fig 8. The same is shown for IEEE 802.11g in Fig 9.

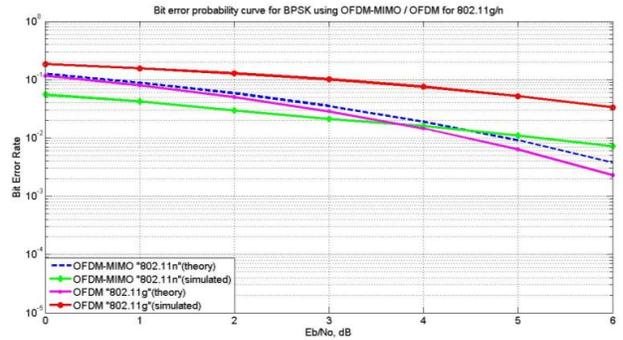

Fig. 7. Simulation results for OFDM (IEEE 802.11g) and OFDM-MIMO (IEEE 802.11n)

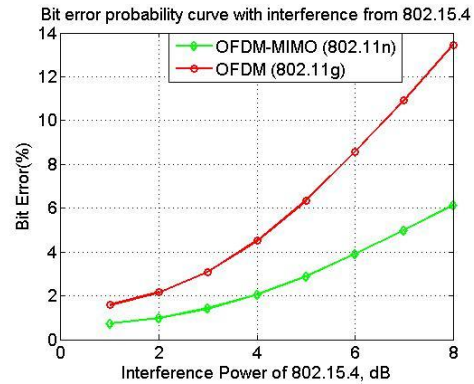

Fig. 8. Simulation results for IEEE 802.11n and IEEE 802.11g in presence of interference from IEEE 802.15.4

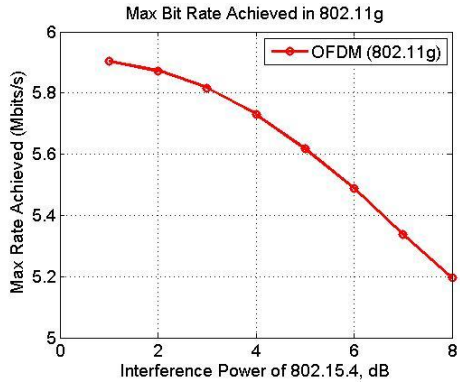

Fig. 9. Maximum throughput for IEEE 802.11g

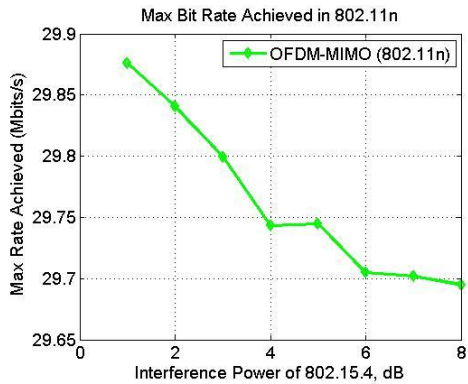

Fig. 10. Maximum throughput for IEEE 802.11n

## 5. Conclusion

The results above confirm that the performance of 802.11n is better than 802.11g in the presence of interference from other sources such as IEEE 802.15.4 in the 2.4GHz ISM band. This result is also intuitive as the spatial diversity using MIMO in IEEE 802.11n makes it more robust and increases its probability of a correct detection (proved in Section 3) due to less dependence on channel and noise conditions. The better performance is also because of the ability of IEEE 802.11n to improve its throughput using multiple data streams. As the throughput of IEEE 802.15.4 (250Kbps) is much less than the throughput of IEEE 802.11n (30Mbits/s for BPSK) this means that IEEE 802.11n can pump more data in the spectrum before it encounters interference from IEEE 802.15.4 traffic. Throughput of IEEE 802.11g (6Mbitss for BPSK) is less than IEEE 802.11n and hence more traffic is obstructed by interference from IEEE 802.15.4 traffic.

As a part of future work the interference from IEEE 802.15.4 can be modeled as a Poisson arrival process instead of the current assumption of packet arrival at equal intervals of time. Although the results obtained in such a setup would be similar to the findings in the current approach, this would provide a much more practical representation of the system.